# Hole self-trapping in the $Y_3Al_5O_{12}$ and $Lu_3Al_5O_{12}$ garnet crystals

V. Laguta, M. Buryi, J. Pejchal, V. Babin, M. Nikl

[1]Institute of Physics AS CR, Cukrovarnicka 10, 16253 Prague 6, Czech Republic

**Abstract**

The processes of hole localization in the $Y_3Al_5O_{12}$ and $Lu_3Al_5O_{12}$ single crystals were investigated by electron paramagnetic resonance (EPR) and thermally stimulated luminescence (TSL). It was found that holes created by x-ray irradiation at 77 K are predominantly self-trapped at regular oxygen ions forming $O^-$ hole center. This self-trapped hole (STH) center is thermally stable to about 100 K in both YAG and LuAG crystals. At higher temperatures, thermally liberated holes are retrapped at oxygen ions in the vicinity of an acceptor ion such as $Mg^{2+}$ and $Al_Y$ or $Al_{Lu}$ antisite ion that leads to increase of the thermal stability of the trapped hole to ≈ 150 K. TSL measurements show two composite glow peaks in the temperature range of 77–280 K, the temperature positions of which well correlate with the thermal stability of the $O^-$ centers. The hole thermal ionization energy was determined from a numerical fit of the TSL peaks within the model of second order kinetics. It is in the range of 0.25-0.26 eV for the $O^-$ STH center, and increases to $0.41 - 0.45$ eV for $O^-$ center stabilized by the acceptor. Revealed $O^-$ centers can be attributed to $O^-$ small polarons formed mainly due to the hole stabilization by short-range interaction with the surrounding lattice.

## 1. Introduction

Among many scintillating single crystals, yttrium aluminum garnet $Y_3Al_5O_{12}$ (YAG) [1,2,3] and especially lutetium $Lu_3Al_5O_{12}$ garnet (LuAG) doped by the luminescent $Ce^{3+}$ or $Pr^{3+}$ ions are known as very efficient scintillators due to high density (LuAG), good light yield and mechanical and chemical resistance [4,5,6,7]. These crystals are usually grown by the Czochralski [8,9] or Bridgman [10] techniques. In spite of the favorable scintillation properties, the main demerit of these materials is the presence of slow components in the scintillation decay [11], which appears due to delayed radiative recombination at $Ce^{3+}$ ions. This phenomenon causes serious degradation of the light yield and timing characteristics as more than 50% of light yield is released in these slower decay components which are undoubtedly related to the host traps [11,12,13]. Origin of such traps in many cases can be clarified by electron paramagnetic resonance (EPR) study especially when the EPR measurements are correlated with corresponding thermally stimulated luminescence (TSL) data [5,14].

During last years, systematic effort has been devoted to further improve time characteristics of the garnet-based scintillators by an admixture of gallium and gadolinium [15,16,17,18,19,20,21,22,23], the effect of which was explained by the removal of shallow electron traps due to the shift of the conduction band edge to lower energies [24]. By composition engineering (tailoring), a new family of materials, so-called multicomponent garnet scintillators has been further developed. The balanced admixture of Ga and Gd into YAG and LuAG, with a general chemical formula $(Gd,Lu,Y)_3(Ga,Al)_5O_{12}$:Ce, resulted in suppressed trapping effect and enormously increased light yield above 50000 ph/MeV [25,26]. Such value considerably exceeds those achieved for the best Ce - doped orthosilicate scintillators.

The lattice defects (or color centers) which cause change or modification of optical and scintillation properties of the YAG and LuAG crystals have been previously studied in many publications (see, for instance, review papers [5,27,28]. Among different possible defects, so-called antisite ion defects, i.e. Y or Lu at the Al site and vice versa and oxygen vacancies are the most frequently mentioned intrinsic defects in garnet crystals [29,30,31,32]. They serve as electron traps [33]. However, it should be emphasized that while the $Y_{Al}$ antisite ions were directly detected by NMR as well as electron traps based on the antisite defect were identified by EPR methods in $YAlO_3$ crystals [34,35], no antisite ions were found in $^{27}$Al and $^{89}$Y NMR spectra of YAG and LuAG [36,37,38,39] suggesting that concentration of such defects is, in fact, much lower than that in $YAlO_3$. On the other hand, oxygen vacancies and related traps, for instance, $F^+$ centers in YAG/LuAG were well documented by EPR [40] and even by electron-nuclear double resonance technique [41].

Although the influence of the antisite and oxygen vacancy-related traps on delayed recombination of carriers in YAG/LuAG can be effectively diminished by Ga co-doping (it lowers the conduction-band edge and removes thus electron trapping effect), similar band-gap "engineering" for hole traps suggested by theoretical calculations [24] has not been yet technologically realized. Therefore, identification and further detailed investigation of the hole traps in garnet crystals is of crucial importance for pushing the performance of these materials close to the intrinsic limits.

It is well known that the most common hole-type defect always present in oxide materials is $O^-$ ion, i.e. a hole trapped at the oxygen lattice ion [42,34,43,44,45]. Such defect can usually be created by a hole self-trapping, but further stabilization of a hole at the oxygen ion can be provided by another defect nearby making the thermal stability of the hole center in the temperature range from about 20-30 K up to room temperatures. Of course, the number of created $O^-$ ions depends on many factors but, in general, the self-trapping is quite effective mechanism of charge carriers trapping. Because $O^-$ ion is paramagnetic (electronic configuration $2p^5$, electronic spin S=1/2), it can be, in principle, easily detected and subsequently studied in detail by EPR. The thermal release of holes from $O^-$ ions also leads to huge thermally stimulated luminescence (TSL) peaks reported for many materials [14,34,43-45].

In spite of the great importance of the $O^-$ centers in garnet crystals there are only a few publications devoted to EPR of these centers. The first observation of EPR spectral line which could be assigned to $O^-$ ions in YAG was mentioned in [40]. However, the most detailed previous study was performed on LuAG ceramics doped with Ce and Mg ions [46]. Some preliminary data were also reported for LuAG crystals doped with Pr and Mg ions [12,47]. In all these publications a hole type paramagnetic center was identified in EPR spectra created by x-ray [12,47] or light [46] irradiation of samples at 77 K. The g factor of this hole center as well as density functional theory (DFT) modeling [46] suggested that it could be $O^-$ ion, i.e. a hole trapped at the $O^{2-}$ lattice ion, presumably in the vicinity of the Mg impurity ion. While this interpretation seems to be quite correct, no valid experimental arguments were presented so far to clarify the nature and actual structure of this center as well as to show a clear correlation of its thermal stability with TSL peaks.

In the present paper, we report results of comprehensive EPR investigation of $O^-$ centers in both YAG and LuAG crystals either nominally pure or doped by $Mg^{2+}$ ions. The EPR data are compared with TSL data obtained for the same crystals. Our results suggest that indeed the holes generated by irradiation at cryogenic temperatures (T < 100 K) in garnet crystals are predominantly self-trapped at oxygen lattice ions in the form of $O^-$ centers. The paper is organized in the following way: After a short description of the applied experimental methods (Sec. 2), we first report our EPR data on $O^-$ center created by x-ray irradiation at 77 K and show that this center is formed by a hole self-trapping (Sec. 3.1). Then we present data on $O^-$ perturbed center (Sec. 3.2). Section 3.3 reports the thermal stability of $O^-$ centers and its correlation with TSL. Finally, in Sec. 4, we draw conclusions.

**2. Experimental details**

The YAG and LuAG crystals were grown by the micro-pulling-down method with radiofrequency inductive heating [48]. An iridium crucible with the die of 3 mm in diameter was used. The growth was performed under the $N_2$ atmosphere using <111> and <100> oriented YAG or LuAG single crystal as a seed. The starting materials were prepared by mixing 4N purity $Y_2O_3$ or $Lu_2O_3$ and $Al_2O_3$ powders. $MgCO_3$ of 4N purity was added to the mixture for the Mg-codoped samples. The concentration of Mg impurity varied from 100 up to 3000 atomic ppm. The crystals were in the form of rods with the diameter of 3 mm and length of 2−3 cm.

The EPR spectra were measured using the commercial Bruker X/Q-band E580 ELEXSYS spectrometer at X-band (microwave frequency 9.25−9.5 GHz) within the temperature range 10-290 K. An X-ray tube operating at a voltage and current of 55 kV and 30 mA, respectively, with Co anode (ISO-DEBYEFLEX 3003 Seifert Gmbh.) was used as a source of X-ray irradiation of crystals at the liquid nitrogen (LN) temperature.

The same X-ray irradiation source was used for TSL excitation. The Optistat cryostat, Oxford Instruments, was used to control the sample temperature. After irradiation at 77 K, the TSL run was initiated with the heating rate of 6 K/min. TSL signal was recorded using two registration channels: (i) spectrally unresolved, by TBX detector downstream the monochromator which was set to zero order in spectrofluorimeter TEMPRO, Horiba Jobin Yvon, with the sampling rate 1 s; (ii) spectrally resolved, by cooled CCD detector in the Ocean Optics spectrometer, in this case, the spectra within 200-800 nm were taken every 10 s. For all measurements, the TSL was monitored within 77−500 K.

### 3. Experimental results

#### 3.1. *Self-trapped hole center*

Before X-ray irradiation, all samples showed EPR spectra of the $Fe^{3+}$ and $Yb^{3+}$ ions always present in YAG and LuAG crystals as accidental impurities [49,50]. After X-ray irradiation at the liquid nitrogen temperature, T=77 K, a new spectrum arises which we further on attribute to $O^-$ ion. It consists of a single broad line without distinctively resolved structure (Fig. 1). The spectrum intensity markedly increases if the crystal is doped by $Mg^{2+}$ ions.

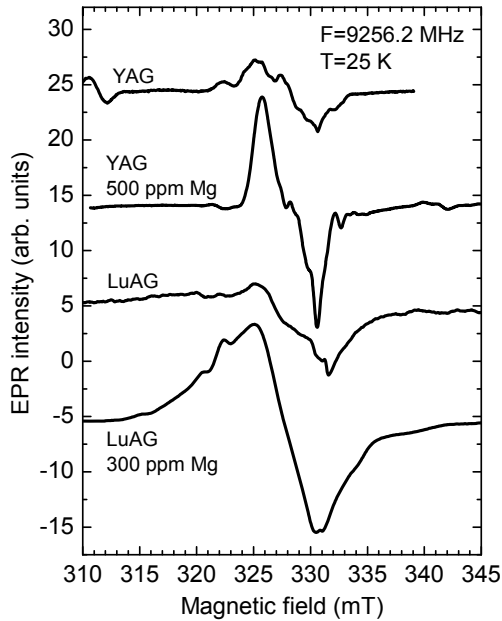

Fig. 1. EPR spectra in YAG and LuAG crystals created by X-ray irradiation at 77 K. The crystals have arbitrary orientation.

The spectral line in LuAG crystals being much broader than that in YAG is completely isotropic, i.e. it conserves both the shape and resonance field under crystal rotation with respect to the magnetic field direction. In contrast to LuAG, the spectral line in YAG being narrower clearly changes its shape under crystal rotation (Fig. 2, left panel) conserving, however, the center of gravity. The specific lineshape with sharp edges suggests that it originates from an unresolved hyperfine (HF) structure. It is confirmed by simulation of the spectra by introducing the HF interaction of electron spin with two Al nuclei having nuclear spin 5/2 and 100% natural abundance. All major variations in the spectral line shape can be obtained by changing HF constant values for two Al nuclei. As an example, the simulated spectrum is shown in Fig. 2, left panel for two crystal orientations (solid blue lines). The simulation was done with the following spin Hamiltonian:

$$H = g\beta SB + I_1 A_1 S + I_2 A_2 S, \qquad (1)$$

where $g = 2.020$ is the g factor, $S = 1$ is the electron spin, $\beta$ Bohr magneton, $B$ magnetic field, $I_1$ ($I_2$) and $A_1$ ($A_2$) are the $^{27}$Al nuclear spins and HF constants, respectively. Depending on crystal orientation the $^{27}$Al HF splitting is in the range of $(3 - 7.7) \times 10^{-4}$ cm$^{-1}$.

Both the positive g factor shift with respect to free electron value and presence of two Al ions in the vicinity of paramagnetic particle indicate that the EPR spectrum belongs to a hole center created by localization of a hole at oxygen in the form of O$^-$ ion. Namely, oxygen sites in the garnet structure have two adjacent Al ions at the distances (LuAG) 1.76 and 1.94 Å [51] as it is seen from the fragment of the YAG (LuAG) structure with a model of the O$^-$ center (Fig. 2, right panel). These two Al ions belong to oxygen tetrahedron (Al$^{tet}$) and oxygen octahedron (Al$^{oct}$).

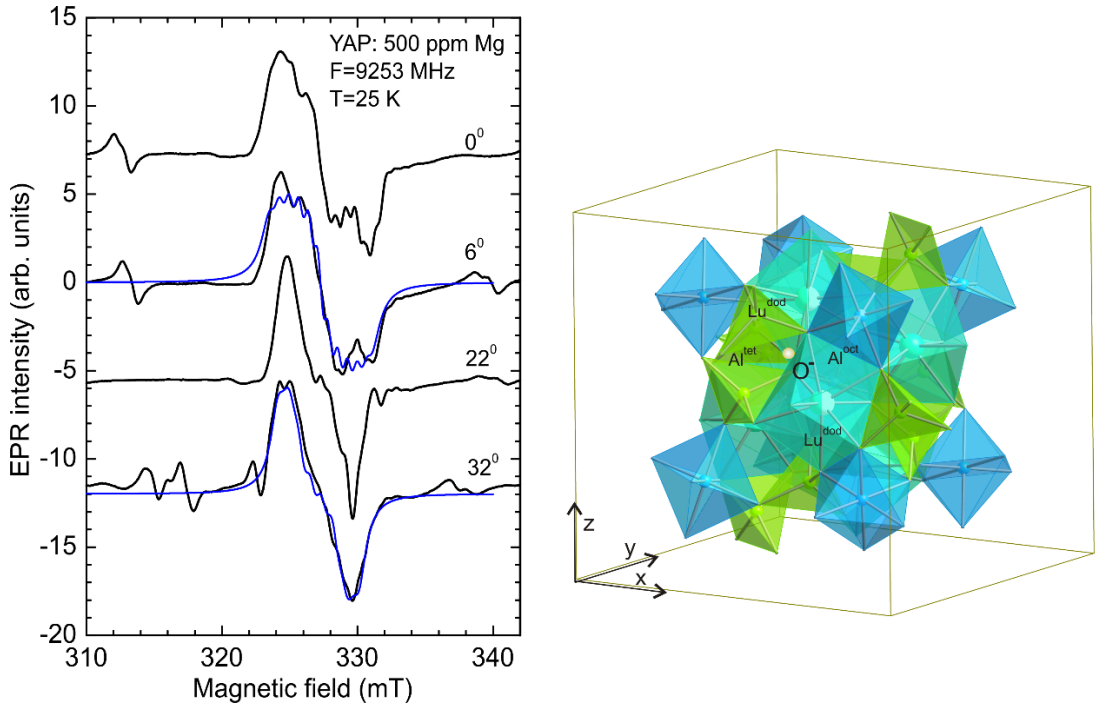

Fig. 2. Left panel: angle variation of the O$^-$ EPR spectrum in YAG: 500 ppm Mg crystal and its simulation (blue solid lines). The right panel shows a fragment of the LuAG crystal structure (unit cell) with O$^-$ ion which is shared by the Al tetrahedron (green color), Al octahedron (blue color) and two Lu dodecahedrons (light blue color).

Of course, the O$^-$ center has several magnetically inequivalent positions in the lattice allowed by the $Ia\bar{3}d$ (230) space group and should also show an anisotropy of the *g* factor. However, all these features are masked by strong HF interactions, which are even much stronger in LuAG leading to Gaussian smooth line shape due to additional HF interaction with nuclear spins of two adjacent Lu ions, Lu$^{dod}$ in Fig. 2. $^{175}$Lu isotope has the nuclear spin of 7/2, natural abundance 97.4% and large both nuclear magnetic and quadrupole moments [52]. Therefore, the spectral line in LuAG is expected to be much broader compared to that of YAG where the $^{89}$Y isotope has a nuclear spin of only 1/2 and more than two times smaller Larmor frequency.

O$^-$ center in both YAG and LuAG is thermally stable to only approximately 100 K. Intensity of its EPR spectrum decays if the crystal is heated to a higher temperature and completely disappears at annealing to 130–150 K (Fig. 3). We assign this O$^-$ spectrum to the self-trapped hole (STH) at O$^{2-}$ ion since this spectrum is present in both YAG and LuAG independently of impurity content in the crystals (see also Fig 1.). The fact that Mg$^{2+}$ doping essentially increases the concentration of these O$^-$ hole centers also supports the model of this center because Mg$^{2+}$ substituting for Y$^{3+}$ or Lu$^{3+}$ acts as an acceptor. It is worth noting that O$^-$ STH center was already identified in many oxide crystals, including YAlO$_3$, see e.g. [34], however, to the best of our knowledge, it has not been yet reported for garnet crystals.

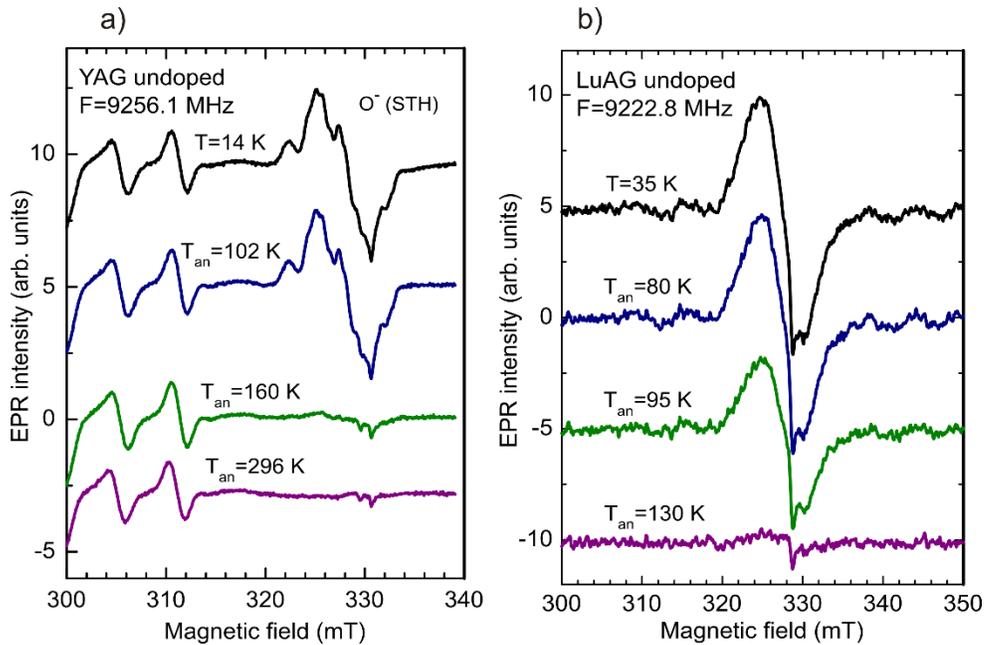

Fig. 3. Dependence of the O$^-$ EPR spectrum in YAG (a) and LuAG (b) on pulsed annealing. X-ray irradiated crystals were heated to a given annealing temperature $T_{an}$, held at that temperature for 4 minutes and then cooled down to the fixed temperature of 14 K or 35 K where the EPR spectrum was measured.

*3.2. O⁻ perturbed center*

When crystals doped by acceptor $Mg^{2+}$ ions are annealed, instead of the STH spectrum, a new spectrum arises. The spectrum is well visible in YAG but due to broad linewidth barely perceptible difference is seen between two spectra in LuAG (Fig. 4). This new spectrum has smaller *g* factor shift (*g* = 2.014–2.015) as compared to the STH spectrum and it is thermally stable almost to room temperature. We attribute this spectrum to a hole trapped at $O^{2-}$ ion perturbed by a defect, most probably $Mg^{2+}$ impurity ion, since its appearance clearly correlates with the Mg doping.

Note that EPR spectrum of x-ray irradiated LuAG:3000 ppm Mg also contains a shoulder at the right side of the spectral line at the magnetic field corresponding to electron-type paramagnetic particle (*g* = 1.975, *g* factor shift is negative). The similar shoulder is also seen in the spectrum of YAG:500ppm Mg crystal (Fig. 4a, the uppermost spectrum). We can thus assume that this signal originates from the trapped electron (TE) center, for instance, $Lu_{Al}$ and $Y_{Al}$ antisite ion which has trapped an electron. The shoulder in the spectrum disappears at almost the same temperature to that when holes liberate from the STH centers indicating that part of the liberated holes recombines at the TE center. Alternatively, the signal at *g* = 1.975 can be ascribed to $F^+$ center, i.e. an electron trapped at oxygen vacancy. But, even in this case, the trapped electron will be delocalized over neighboring cations. The *g* factor shift will mainly be determined by a contribution from $d^1$ orbitals of Lu or Y ions similar as, for instance, in $F^+$ center in YAP [34,14] and $Y_2SiO_5$ [44].

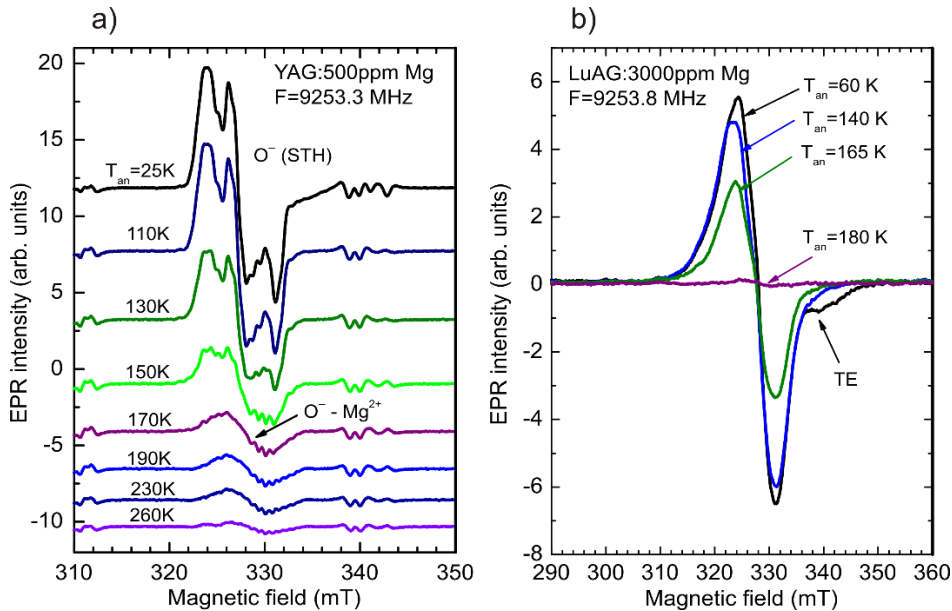

Fig. 4. Dependence of the O⁻ EPR spectrum in YAG (a) and LuAG (b) on pulsed annealing showing the transformation of the O⁻ STH spectrum into the spectrum of O⁻ ion perturbed by a defect nearby. The shoulder at the right side of the spectral line originates from the trapped electron center.

In YAG crystals the new O⁻ spectrum shows partly resolved HF structure, illustrated in Fig. 5 together with the simulated HF structure.

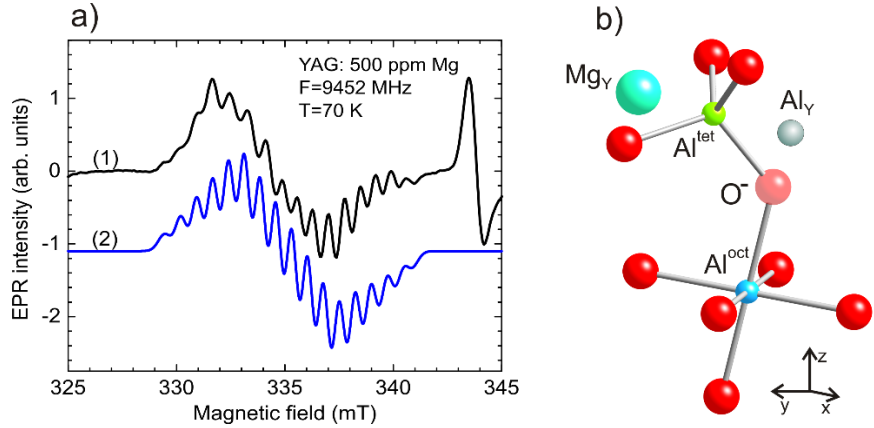

Fig. 5. (a) EPR spectrum of the O⁻ perturbed center (1) and its simulation (2) in YAG doped with 500 ppm Mg. (b) Model of the O⁻ perturbed center: it is assumed that the perturbation is due to the presence of $Mg^{2+}$ impurity and $Al_Y$ antisite ion.

The simulated HF structure was calculated using the spin Hamiltonian (1) with three HF terms:

$$H = g\beta SB + I_1 A_1 S + I_2 A_2 S + I_3 A_3 S. \qquad (2)$$

The simulated parameters are: $g = 2.015$; $I_1 = I_2 = I_3 = 5/2$; $A_1 = 7.3 \times 10^{-4}$ cm$^{-1}$, $A_2 = 6.8 \times 10^{-4}$ cm$^{-1}$ and $A_3 = 6.3 \times 10^{-4}$ cm$^{-1}$ with each HF component having the same Gaussian linewidth of $6.7 \times 10^{-4}$ cm$^{-1}$. The fit is not ideal. However, the simulated spectrum properly describes all characteristic features of the measured one such as the number of HF components and the distribution of their intensities. The discrepancy between calculated and measured spectrum is related to slight inequality of HF splitting along the HF pattern due to anisotropy of both the $g$ factor and HF interaction. It was difficult to include this anisotropy in the calculation due to its uncertain value and existence of several magnetically inequivalent positions of the center.

The HF parameters determined suggest the presence of third Al ion in the vicinity of O⁻ ion. In regular garnet lattice, except for two adjacent Al ions, all other Al ions are located far from oxygen ions (the shortest next-nearest-neighbor O − Al distance is 3.297 Å). Then, the only reasonable possibility to explain the HF structure is the presence of Al ion at the Lu site, so-called $Al_{Lu}$ antisite defect, widely discussed in literature [29-33]. In this case, the above determined HF parameters are well explained as due to the interaction of hole with nuclear magnetic moments of three $^{27}$Al nuclei located at distances of 1.76, 1.94 and 2.276 Å from the oxygen ion in the unrelaxed lattice (Fig. 5b). In this model of the center, we cannot be completely sure if the second closest $Lu^{3+}$ ion is substituted by $Mg^{2+}$ ion. Mg has one isotope with the non-zero nuclear spin $I = 5/2$, but it has only 10% natural abundance that makes its contribution to HF structure negligibly small. From a general point of view and in accordance with the density functional theory (DFT) calculation [46], the $O^- - Mg^{2+}$ charge compensated pair is energetically favorable. Additional stabilization of the center is provided by $Al_{Lu}$ or $Al_Y$ antisite ion. However, while the antisite ion changes spectral parameters of $O^-$ ion, mainly HF interaction, its influence on the thermal stability of trapped hole should be much weaker as compared to $Mg^{2+}$ acceptor ion. Such $O^-$ centers stabilized by an acceptor are the most frequently occurring defects created under light or x-ray irradiation in oxides [42]. Note that similar $O^-$ centers have been observed in $YAlO_3$ [53,34]. The detailed study showed that they are attributed to small (or bound small) polarons [53]. In principle, the modern theories of polarons consider the self-trapped particle as an extreme case of small polarons (localized small polarons) [54,55] that explains the mechanism of particle localization in the lattice as due to strong electron-phonon interaction. In highly symmetric lattices the localized electron or hole state can be additionally stabilized by the Jahn-Teller or pseudo Jahn-Teller local distortion of lattice [56]. Such polaron is called Jahn-Teller small polaron. Obviously, the Jahn-Teller mechanism could be realized in YAG/LuAG lattice due to its cubic symmetry.

### 3.3. Thermal stability of $O^-$ centers and its correlation with TSL

The thermal stability of the X-ray created $O^-$ centers was studied by the pulse annealing method. After irradiation at 77 K, the sample was quickly cooled down to 20–30 K and then it was heated with a rate of 0.5-1 K/s up to a certain temperature $T_{an}$, held at that temperature for 4 minutes, and then quickly cooled down (with a rate of 1-2 K/s) to a fixed temperature 25 K where the EPR intensity was measured. The measurements at a low temperature were necessary in order to avoid the influence of the spin-lattice relaxation on the signal intensity. Except for the irradiation, which was carried out only once during the first step, this procedure was repeated for different temperatures $T_{an}$. The 4 minute interval was determined as the optimum balance between good thermalization of the sample and sufficient reproducibility of the measured EPR intensities. However, due to limitation in the heating and cooling speeds the actual annealing time was about 20-25% longer. The spectrum intensities obtained by double integration of spectral line were recalculated in relative concentrations for $O^-$ centers and are depicted in Fig. 6 for YAG:Mg and LuAG:Mg. It can be seen that the concentration of the $O^-$ STH centers diminishes at the annealing to 120–130 K in YAG. Approximately at these temperatures the $O^- - Mg^{2+}$ center appears. Its concentration decreases already at T > 190 K. This center completely disappears at annealing temperature of approximately 250 K. In case of LuAG crystals, spectral lines from $O^-$ STH center and $O^-$ center stabilized by an impurity ion can hardly be distinguished apart. Therefore, experimental data for LuAG in Fig. 6 are presented as a sum of contributions from both types of the centers. Nevertheless, one can see marked decrease in the total concentration of O⁻ centers at the temperature where STH centers disappear in YAG. It confirms that similar O⁻ STH centers exist in LuAG as well. The further sharp decrease of the O⁻ concentration corresponds to the thermal disintegration of O⁻ – $Mg^{2+}$ centers.

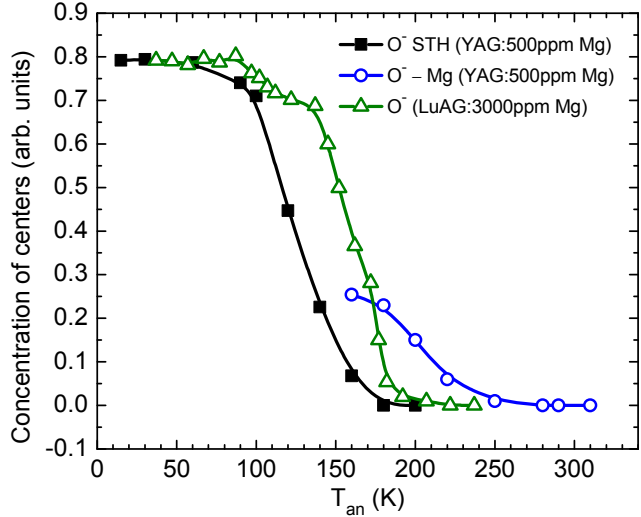

Fig. 6. Relative concentrations of O⁻ centers in YAG: 500 ppm Mg and LuAG: 3000 ppm Mg crystals (symbols) as a function of pulse annealing temperature.

In order to get a deeper insight into the kinetics of the trapped holes and to determine their thermal stability parameters, we have performed TSL measurements of the crystals irradiated by x-ray at 77 K. The results are shown in Fig. 7 for the same crystals used in EPR measurements. The first intense TSL peak in both YAG and LuAG is detected at 118-122 K, exactly at the temperatures where O⁻ STH center undergoes thermal destruction. The second, composite TSL peak around 200 K is obviously related to the thermal disintegration of the O⁻ – $Mg^{2+}$ centers. Intensity of this peak substantially diminishes in undoped crystals and completely disappears in high purity crystal (see, e.g. Fig. 2 in [13]). The peaks at T > 280 K may originate from the thermal release of electrons from oxygen vacancies or $Lu_{Al}$ ($Y_{Al}$) antisite defects. However, their origin needs further investigation.

The TSL glow curve numerical analysis was based on the well-known formula of the Garlick-Gibson [57] valid for the second order recombination kinetics. This kinetics leads to almost symmetrical shape of TSL glow curve in agreement with the experiment. The results of the TSL peaks simulation are displayed in Fig. 7 by the red and grey solid lines to the temperature 260 K. The fit is quite good taking into account the composite character of the TSL peaks. Nevertheless in the simulation of TSL curve, we tried to use a minimum number of traps to provide only characteristic features of the observed TSL. In particular, we assumed that only one trap (O⁻ STH center) mainly contributes to the first peak at 118-122 K. The second, composite TSL peak around 200 K was simulated assuming the presence of two traps contributing to this peak (grey lines in Fig. 7). Thermal stability parameters of the traps are listed in Table I. One can see that the peaks related to the O⁻ STH center is described by approximately the same parameters in all three crystals: the thermal depth is 0.252-0.264 eV and the effective frequency factor $s'$ varies from $4.6\times10^6$ s⁻¹ m³ in YAG to $5\times10^7$ s⁻¹ m³ in LuAG. The second two peaks at 186-214 K have much larger thermal depth of 0.409-0.45 eV. The complex character of the TSL curve around 200 K with variation in shape from crystal to crystal suggests the presence of several traps of the same origin, namely originating from O⁻ hole centers stabilized by a defect nearby. We also cannot exclude more complex character of the hole excitation from O⁻ ion than the simple one ascribed by the Arrhenius law. In any case, it is clear that the main contribution to TSL in Mg-doped crystals around 200 K is coming from the $O^- - Mg^{2+}$ centers as this peak substantially diminishes in undoped pure crystals.

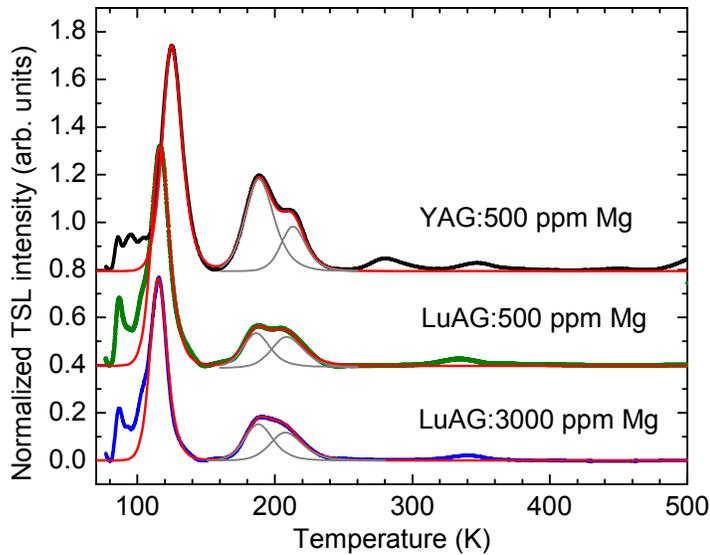

Fig. 7. Normalized TSL glow curves of YAG:500 ppm Mg, LuAG:500 ppm Mg and LuAG:3000 ppm Mg crystals irradiated by x-rays at 77 K measured at the heating rate of 6 K/min. Red solid lines are simulated TSL peaks; grey lines represent individual peaks for the composite TSL peak at 200 K. Glow curves are vertically shifted for better visualization.

Table I. Thermal stability of traps visible in TSL glow peaks in YAG and LuAG crystals with Mg impurity. The error in $E_T$ value is less than 5%, while it could be up to 25% for the effective frequency factor $s'$.

| Crystal | $T_m$ (K) | Thermal depth $E_T$ (eV) | Frequency factor $s'$ (s$^{-1}$ m$^3$) |
|---|---|---|---|
| YAG:500ppm Mg | 125 | 0.264 | $4.6\times10^6$ |
| | 188 | 0.409 | $1\times10^7$ |
| | 214 | 0.62 | $1.5\times10^{11}$ |
| LuAG:500ppm Mg | 116 | 0.252 | $1\times10^7$ |
| | 186 | 0.44 | $3\times10^8$ |
| | 209 | 0.45 | $2\times10^7$ |
| LuAG:3000ppm Mg | 116 | 0.262 | $5\times10^7$ |
| | 188 | 0.418 | $5\times10^7$ |
| | 208 | 0.446 | $2\times10^7$ |

## 4. Conclusions

EPR measurements were carried out in YAG and LuAG crystals irradiated by x-rays at 77 K. The crystals were both nominally pure and doped by Mg. The measurements clearly revealed O$^-$ hole centers which were attributed to (i) self-trapped hole at the lattice oxygen ion and (ii) a hole at oxygen ion additionally stabilized by a perturbing defect nearby, namely, in the case of Mg doping, this defect is Mg$^{2+}$ substituting for Y$^{3+}$ or Lu$^{3+}$ and Al$_Y$ or Al$_{Lu}$ antisite ions.

The O$^-$ STH center is thermally stable to about 100 K in both the YAG and LuAG crystals. The O$^-$ perturbed center (O$^-$ − Mg$^{2+}$) is stable to ≈150 K. These hole centers can be attributed to O$^-$ small polarons. In the small polaron model, the hole localization is caused predominantly by the short-range lattice distortions. Therefore, more appropriate approach to model such centers should be based on the DFT calculation as it was demonstrated in [46]. In particular, the trap depth of 1 eV was predicted for the Mg-perturbed O$^-$ center in LuAG. This energy is only about two times larger than the measured one demonstrating quite good result as for the first principles calculation.

Besides the hole-type centers, x-ray irradiation also creates trapped electron center characterized by the *g* factor 1.975. However, no detailed studies were performed for this center.

TSL measurements performed for the same crystals used in EPR show approximately the same composite glow peaks in YAG and LuAG crystals in the temperature range of 77–280 K. The temperature positions of the peaks well correlate with thermal stability of the O$^-$ STH and O$^-$ − Mg$^{2+}$ centers suggesting that these peaks mainly originate from thermal destruction of the O$^-$ centers.

Thermal stability parameters of the traps contributing to TSL (trap depth/ionization energy and frequency parameter) were determined from numerical analysis of corresponding peaks based on the formula of Garlick-Gibson [57]. In particular, the ionization energy of the $O^-$ STH center is 0.25-0.26 eV. It increases to 0.41-0.45 eV for the $O^- - Mg^{2+}$ centers.

**Acknowledgements:** The financial supports of the Ministry of Education, Youth and Sports of Czech Republic, projects SAFMAT LM 2015088 and LO1409 and Czech Science Foundation grant 17-09933S are gratefully acknowledged.